\documentclass[conference]{IEEEtran}
\ifCLASSOPTIONcompsoc
  \usepackage[nocompress]{cite}
\else
  \usepackage{cite}
\fi
\ifCLASSINFOpdf
  \usepackage[pdftex]{graphicx}
  \graphicspath{{figures/}}
  \DeclareGraphicsExtensions{.pdf,.jpeg,.png,.svg}
  \usepackage{balance}

\else
\fi
\usepackage{amsmath}

\usepackage{color}

\definecolor{ForestGreen}{RGB}{5,176,5}

\usepackage[hyphens]{url}
\usepackage[bookmarks=false]{hyperref}

\hypersetup{breaklinks=true}
\hypersetup{hidelinks}
\usepackage{cite}

\usepackage{balance}
\usepackage{multirow}
\usepackage{tikz}

\begin{document}
%
\title{Using Undersampling with Ensemble Learning to Identify Factors Contributing to Preterm Birth}

\author{
	\IEEEauthorblockN{Shi Dong\IEEEauthorrefmark{1}, Zlatan Feric\IEEEauthorrefmark{1}, Guangyu Li\IEEEauthorrefmark{2}, Chieh Wu\IEEEauthorrefmark{1}, April Z. Gu\IEEEauthorrefmark{3}, Jennifer Dy\IEEEauthorrefmark{1}, John Meeker\IEEEauthorrefmark{5},\\ Ingrid Y. Padilla\IEEEauthorrefmark{6}, Jose Cordero\IEEEauthorrefmark{7}, Carmen Velez Vega\IEEEauthorrefmark{8}, Zaira Rosario\IEEEauthorrefmark{6}, Akram Alshawabkeh\IEEEauthorrefmark{2}, David Kaeli\IEEEauthorrefmark{1}}  
	\IEEEauthorblockA{\IEEEauthorrefmark{1}Dept. of Electrical and Computer Engineering, Northeastern University}
	\IEEEauthorblockA{\IEEEauthorrefmark{2}Dept. of Civil and Environmental Engineering, Northeastern University}
	\IEEEauthorblockA{\IEEEauthorrefmark{3}Department of Civil and Environmental Engineering, Cornell University}
	\IEEEauthorblockA{\IEEEauthorrefmark{5}University of Michigan  \IEEEauthorrefmark{6}University of Puerto Rico at Mayaguez  \IEEEauthorrefmark{7}University of Georgia}
	\IEEEauthorblockA{\IEEEauthorrefmark{8}Graduate School of Public Health, University of Puerto Rico Medical Campus}
}
\maketitle

\thispagestyle{empty}
\pagestyle{empty}

\begin{abstract}

In this paper, we propose Ensemble Learning models to identify factors contributing to preterm birth. Our work leverages a rich dataset collected by a NIEHS P42 Center that is trying to identify the dominant factors responsible for the high rate of premature births in northern Puerto Rico. We investigate analytical models addressing two major
challenges present in the dataset: 1) the significant amount of incomplete data
in the dataset, and 2) class imbalance in the dataset. First, we leverage and compare two types of missing data imputation methods: 1) mean-based and 2) similarity-based, increasing the completeness of this dataset. Second, we propose a feature selection and evaluation model based on using undersampling with Ensemble Learning to address class imbalance present in the dataset. We leverage and compare multiple Ensemble Feature selection methods, including Complete Linear Aggregation (CLA), Weighted Mean Aggregation (WMA), Feature Occurrence Frequency (OFA) and Classification Accuracy Based Aggregation (CAA). To further address missing data present in each feature, we propose two novel methods: 1) Missing Data Rate and Accuracy Based Aggregation (MAA), and 2) Entropy and Accuracy Based Aggregation (EAA). Both proposed models balance the degree of data variance introduced by the missing data handling during
the feature selection process, while maintaining model performance. Our results show a 42\% improvement in sensitivity versus fallout over previous state-of-the-art methods.

\end{abstract}


%
\IEEEpeerreviewmaketitle

\section{Introduction}
Machine Learning (ML) techniques have shown significant potential in facilitating public health research. Unlike conventional data-driven statistics and analytics in the public health domain, ML can discover hidden patterns and knowledge in large amounts of data, significantly facilitating the research process. In recent years, public health research institutes have invested a vast amount of effort on: i) increasing the availability of research data, and ii) investigating and building accurate and reliable analytical models with ML and AI algorithms~\cite{CHEN2016688}.


Although ML algorithms provide a rich collection of candidate methods, building a reliable ML-based analytical model can be challenging. For instance, ML algorithms provide a range of performance and accuracy based on the characteristics of the input dataset. Selecting the right ML algorithm requires a thorough understanding of the characteristics of the given dataset. After building the analytical model, we can apply the same model to all datasets with similar characteristics. 

We have developed an analytical model that can identify the key factors tied to preterm birth. Our work is informed through access to a rich dataset from an National Institute for Environmental Health Sciences (NIEHS) P42 Center. Unfortunately, analyzing this dataset presents two major challenges. First, given the nature of the data collected, we encounter a significant number of missing data entries in a variety of fields, resulting in incomplete records. Second, given that the dataset includes a limited number of preterm birth outcomes (anomalies), the dataset is highly imbalanced in terms of class distribution, where the number of negative samples (term births) is several times larger than the number of positive samples (preterm births). If we attempt to use standard algorithms on this data, the outcome imbalance will clearly bias our results, misleading researchers, and resulting in poor model performance.

In this paper, we explore analytical models that address these two challenges present in the dataset. First, to impute the missing data, we leverage and compare two types of missing data imputation methods, i.e., mean-based and similarity-based algorithms. Next, we propose a model built using undersampling with ensemble learning~\cite{easyensemble}, incorporating Decision Trees~\cite{flach} and Recursive Feature Elimination (RFE)~\cite{sklearn} to rank and select features. We also leverage multiple commonly used ensemble feature selection methods, including Complete Linear Aggregation (CLA), Weighted Mean Aggregation (WMA), Feature Occurrence Frequency (OFA), and Classification Accuracy Based Aggregation (CAA)~\cite{Brahim2018EnsembleFS}. As a further step, to address the high variance introduced by missing data handling, we propose two novel methods: 1) {\em Missing Data Rate and Accuracy Based Aggregation (MAA)} and 2) {\em Entropy and Accuracy Based Aggregation (EAA)}, which are both based on CAA. The former leverages the missing data rate of each feature, selecting features that have a lower degree of missing data. The latter leverages the information entropy change of each feature after applying missing data handling, selecting features that preserve the original entropy. We evaluate and compare these two methods, and provide insights when leveraging them during feature selection for the dataset. We use both accuracy and Area Under the ROC Curve (AUC) metrics to evaluate the performance, considering that the data is highly imbalanced in terms of the outcome class distribution. The experiment results show a 42\% improvement in AUC over previous state-of-the-art results on the same data, which pursued a hybrid approach~\cite{dongbigdata}.


\section{Related Work}\label{relatedwork}
The use of Machine Learning analytics in public and environmental health research has
attracted global attention.
Auffray et al. evaluate the potential benefits of leveraging Big Data
techniques for public health in Europe~\cite{Auffray2016}. Chen et al. present
the work of leveraging cognitive computing to address Big Data challenges in
medical research~\cite{CHEN2016688}. Zheng et al. present a study of air
pollution using Big Data techniques~\cite{Zheng2013}. Beam et al. elaborate the
connections between Big Data technologies and health studies, clarifying their
roles in each research domain~\cite{beamjama}.

Other than general work related to Big Data analytics, our work here focuses on
addressing challenges of missing data and outcome class imbalance. Many
previous studies have considered these issues. For missing data imputation,
multiple techniques are available, including mean substitution, multiple imputation and full
information maximum likelihood, and have been applied to survey and
clinical data~\cite{goeij2013, Wu2012, roderick,Sterneb2393,Schlomer2010BestPF}. There are multiple
techniques for tackling class imbalance as well. Prior studies adopted using a Decision
Tree combined with AUC as training criterion to select features with the best
AUC performance~\cite{dongbigdata}. SMOTEBoost~\cite{smoteboost} and
RUSBoost~\cite{rusboost} are two similar approaches leveraging oversampling and
ensemble of weak learners that are commonly used in boosting. Balanced random
forest~\cite{Chen2004UsingRF} and EasyEnsemble~\cite{easyensemble}, on the other
hand, leverage undersampling and ensemble learning, on which our work is based.

Ensemble feature selection methods have long been investigated and applied in
bioinformatics, especially given their robustness when identifying biomarkers. Brahim et al.
analyze and compare multiple commonly used ensemble feature selection
methods~\cite{Brahim2018EnsembleFS}. Abeel et al. propose two aggregation
methods, Complete Linear Aggregation (CLA) and Weighted Mean Aggregation
(WMA)~\cite{Abeel2009RobustBI}. The former aggregates the complete feature rank
lists across each instance of ensemble and selects a set of features with the
highest ranks. The latter uses the mean of the AUC as a weight for ranking. Chan et al.
propose classification Accuracy Based Aggregation (CAA)~\cite{chancca}. 
In this paper, we investigate these various ways for performing feature selection from ensemble methods.  
Furthermore, since our data has a high percentage of missing data entries, we introduced a novel strategy for taking data missing rate into account (via the effect of data imputation) in ensemble feature selection.

The previous work relied on correlation, SVMs and
PCA~\cite{Abeel2009RobustBI, chancca} to select features. In our work we used Decision
Trees because they have shown to produce superior performance on this specific
dataset~\cite{dongbigdata}, and intrinsically performs feature selection.

\section{Background}\label{background}
\subsection{The NIEHS Dataset Targeting Preterm Birth}\label{sec:protect_data}


The NIEHS has established a Center to study environmental links to the high preterm birth rates observed in northern Puerto Rico. Over an 8 year period, the Center has collected both environmental data (e.g., water samples, air samples, product use surveys) and personal health data (e.g., chemicals in blood, urine, and placenta, health records, birth outcomes) for a cohort of over 2000 expectant mothers in this region, as well as the birth outcome (i.e., preterm birth or term birth)~\cite{dongdatabase}. The dataset consists of a wide range of sources, including:

\begin{itemize}
\item Human subjects information - medical history, reproductive health records, product use data surveys, and birth outcomes 
\item Biological samples - blood, urine, hair and placenta samples 
\item Environmental samples and measurements - soil samples, well and tap water samples, historical Environmental Protection Agency (EPA) data, soil samples, superfund site data
\end{itemize}

The Center has made their data available to the broader research community. We have gained access to this large repository of environmental health data records, enabling us to explore analytical models to identify linkages between a large number of potential contributing factors and the high premature birth rate. We consider one subset of data in this paper, i.e., the human subject data. It contains forms capturing demographics and lifestyle information, as well as medical and health records. The dataset includes many key factors that can help guide domain-specific research. Additionally, the collection procedure for the human subject data is carried out in four phases chronically, associated with four visits according to different stages of the mother's pregnancy. The data subset analyzed in this paper represents information collected at different pregnancy stages, defined as visit 1 (V1) to visit (V3), which correspond to the trimesters, and with visit (V4) included as a postpartum collection.

\section{Missing Data Imputation}\label{missingdatahandling}
\subsection{Data Preprocessing Model}\label{sec:datapreprocessing}

Missing data entries pose significant challenges to machine
learning algorithms. We developed a series of tools to preprocess
the data stored in the database, generating valid inputs that can be
readily consumed by the analytical model. Our
tools provide an end-to-end processing pipeline.

\begin{figure}[t!] \centering
\includegraphics[width=\columnwidth]{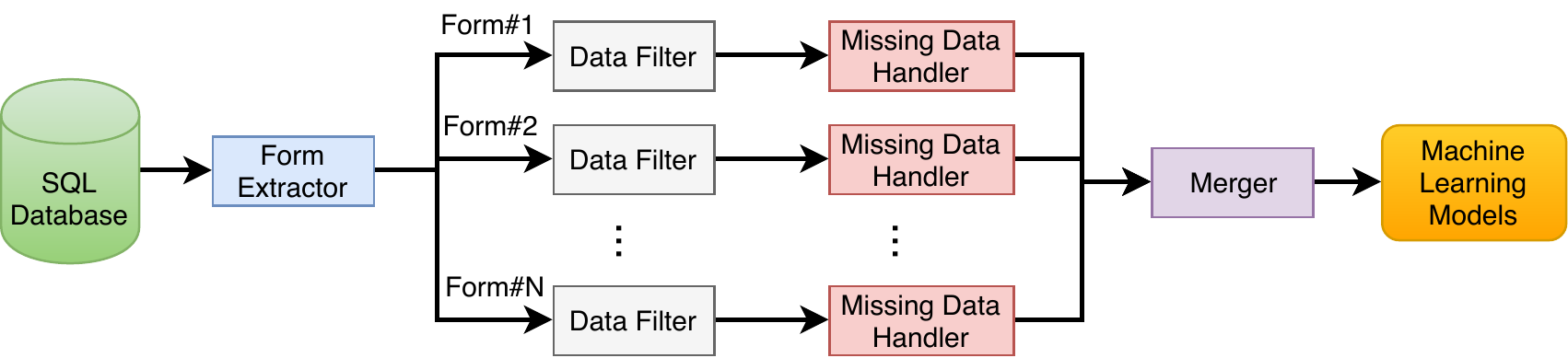} 
\caption{Analytical tool
chain for data preprocessing.} 
\label{fig:endtoend} 
\end{figure}

Specifically, the tool chain consists of components including a form extractor,
data filter, missing data handler and merger.  Figure~\ref{fig:endtoend} shows
the workflow of our tool chain.  First, the form extractor obtains the selected
forms for a set of selected patient IDs which correspond to
pregnant mothers. Second,
data is fed to the data filter and missing data handler. The data filter
takes care of two things: 1) filtering out text-based data entries, 
and 2) filtering
out features that have less than $P\%$ of the participants
completing the field, and
filtering out patients having less than the $P\%$ of their feature values
completed, in which, $P\%$ is a configurable parameter, representing a data
completion rate threshold. After data filtering, each form only contains
numerical and categorical data values, and at least $P\%$ valid data values on both
a feature basis and participant basis. The missing data handler implements the
means-based or similarity-based imputation algorithm.


\subsection{Missing Data Imputation Algorithm}

To address missing data handling, we leverage two commonly-used missing data imputation methods: 1) mean-based imputation and 2) similarity-based imputations~\cite{goeij2013, Wu2012}. Applying mean-based algorithm is straightforward. The missing data entries in each feature are replaced with the mean value of the corresponding feature. For the categorical data, we replace the missing data entries using the category closest to the mean value. But applying a similarity-based imputation is challenging on the given dataset. The dataset contains mixed data types (i.e., textual, numerical and categorical data) , restricting the use of standard data imputation methods which are commonly used when targeting a specific type of data. Alternatively, we propose a customized algorithm that can handle both categorical and numerical data, based on K-nearest Neighbor Imputation~\cite{Kim2004}. To calculate the similarity, we use the following equation:

\begin{equation}\label{eq:similarity} S_{ij} = \frac{\vec{x_{i}} \cdot
\vec{y_{j}}}{N} 
\end{equation} 
where \begin{math}S_{ij}\end{math} stands
for the similarity between sample vectors \begin{math}x_{i}\end{math} and
\begin{math}y_{j}\end{math}.  The \begin{math}\cdot\end{math} denotes a dot
product operation and \begin{math}N\end{math} denotes the number of features
involved in the dot product operation. Note that the numerical data is normalized and the categorical data is transformed to a one-hot encoding format.

\section{Undersampling Ensemble Feature Selection}\label{ensemblemethod}

\subsection{Undersampling Ensemble Learning}

Undersampling forcibly balances positive and negative samples to address the
class imbalance issue in a straightforward manner. In a binary class problem, an
instance of undersampled training data can be constructed with \textit{\textbf{equal-sized}}
samples from both the majority and minority classes.
For example, one dataset contains two classes, positive and
negative outcomes. The positive samples 
are assigned to subset $P$ and the negative samples are assigned to
subset $N$. When the class distribution is highly imbalanced, the size of $P$
($|P|$) is several times smaller than the size of $N$ ($|N|$). The undersampling
randomly selects samples from $N$ to construct a subset $\bar{N}$ so that
$|P|=|\bar{N}|$, and then combines them into a single 
training data instance, i.e.,
$P\bigcup\bar{N}$.

Although the new instance of 
the training data is balanced, we lose a lot of 
information from the majority class. Solely relying on this training data can
lead to poor model performance. To deal with this problem, a form of ensemble
learning can be applied using multiple instances 
of training data in a set, i.e.
$\{P\bigcup\bar{N}_0, P\bigcup\bar{N}_1,...,P\bigcup\bar{N}_{n-1}\}$, in which
each $\bar{N}_i$ is a subset of 
the samples from the majority class obtained through
undersampling, where $n$ is the number of ways.

In our proposed model, which is 
based on undersampling with ensemble
learning, we elect to incorporate Decision Trees~\cite{flach} as the base classifier and
feature selector, considering that it can
serve as both a feature ranker and a classifier.

\begin{figure}[t!] \centering
\includegraphics[width=\columnwidth]{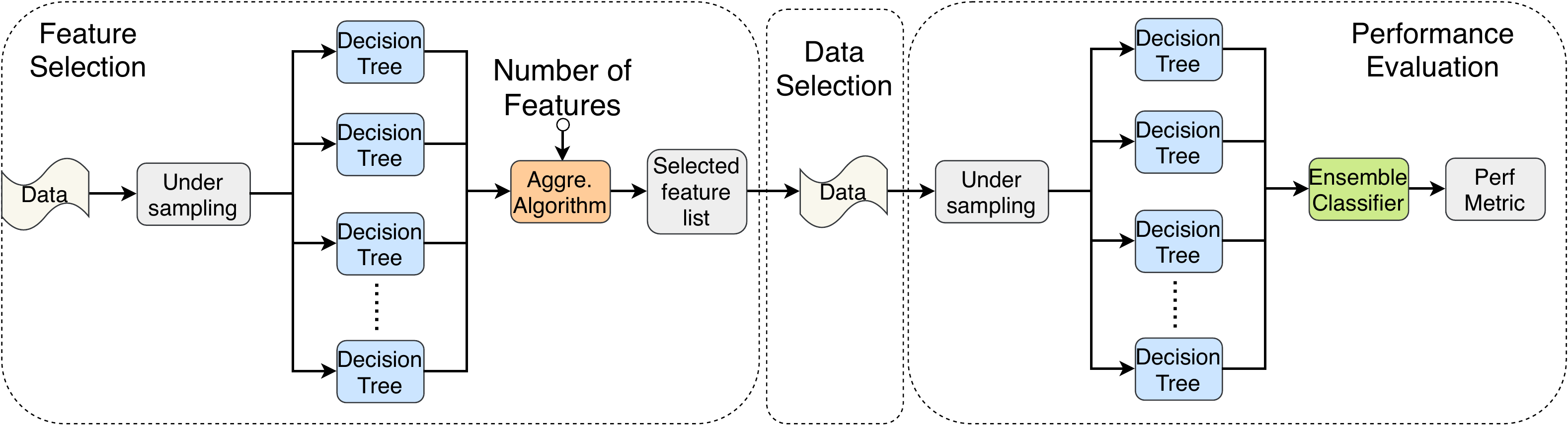}
\caption{Undersampling ensemble learning model for feature selection and
evaluation.} 
\label{fig:ensemblefs} 
\end{figure}

Figure~\ref{fig:ensemblefs} shows the proposed analytical model leveraging
undersampling ensemble learning with Decision Trees. 
The model consists of three
steps: i) feature selection, ii) data selection and iii) performance evaluation.
In the feature selection step, an ensemble of training data 
sets are produced by applying the
undersampling procedure described above. 
Each training set will be used to build a Decision
Tree to rank the entire feature set during the training process. Then  
an ensemble feature selection method (aggregation algorithm) is applied to
produce the final
selected feature set, according to a configurable number of features. We will
explore a collection of commonly used ensemble feature selection methods in the
next section. 
In the
data selection step, a subset of data is selected according to
the selected features. During performance evaluation, we follow a similar
process. The only difference is that the
Decision Tree built using each training data instance is used
for prediction
instead of ranking features. 
The classification results from each ensemble
instance will be fed to an ensemble classifier to make a final prediction. In
this work, we use a simple majority voting as the ensemble classifier. Based on
the prediction, we calculate performance metrics such as accuracy and AUC.

\subsection{Ensemble Feature Selection}


In this section,
we describe four commonly-used ensemble feature selection methods used as the
aggregation algorithms shown in figure~\ref{fig:ensemblefs}.

Complete Linear Aggregation (CLA) and Weighted Mean Aggregation
(WMA)~\cite{Abeel2009RobustBI} are representative of the
class of algorithms that perform aggregation
of list rankings, using the
numbers ranging from 1 to the number of total features. 
Here, 1 represents the highest rank. CLA simply sums up
every list ranking to obtain the final 
ranking results for every feature. Finally,
a number of features with highest ranks (smallest scores) are selected. WMA is
an advanced version of CLA, treating model performance metrics of each feature
selector as weights applied to
feature ranks. Before summing up ranks, each rank will
be weighted by a performance metric from the associated feature selector. 

In our
proposed model, Recursive Feature Elimination (RFE) is used to generate ranks
based on the feature ratings from the Sklearn Decision Tree~\cite{sklearn}. For
WMA, we use $1-AUC$ as the weight for each rank. The $AUC$ is obtained through a
10-fold cross validation using the training data for each ensemble instance.

Feature Occurrence Frequency (OFA) and Classification Accuracy Based Aggregation
(CAA)~\cite{Brahim2018EnsembleFS} represent a class of algorithms that 
count the
occurrence of each selected feature. 
OFA is the most straightforward method, only
counting the occurrence of selected features from each ensemble instance. CAA
takes one step further, weighing the counts for each 
occurrence with an associated classification
accuracy. When selecting features, the ones with the largest score
(weighted count of occurrence in CAA) are selected. In our proposed model, for
each feature selector, we select features with only positive ratings 
using a Decision
Tree and validate the AUC and accuracy using 10-fold cross validation.

\subsection{Missing Data Rate/Entropy And Accuracy Based Aggregation (MAA/EAA)}

Next, we present two novel ensemble-based feature selection methods:
1) Missing Data Rate And Accuracy Based Aggregation (MAA)
and 2) Entropy And Accuracy Based Aggregation (EAA). Both methods are based on
CAA, using the frequency of occurrence as the score. In addition to leveraging 
classification accuracy, our proposed approaches also consider mitigating the 
impact of the potential increase in variance introduced by data imputation when ranking features. 
MAA uses the missing data rate of each feature to adjust the weight, whereas EAA 
uses the change in entropy for each feature, computed before and after applying missing data handling.
Equations~\ref{maa} and \ref{eaa} show the adjusted weights for MAA and EAA, respectively.

\begin{equation}\label{maa} 
\frac{Accuracy}{(Missing\_rate+\alpha)^\beta}
\end{equation}

\begin{equation}\label{eaa} 
\frac{Accuracy}{(\bigtriangleup Entropy+\alpha)^\beta}
\end{equation}

In the above two equations, $Missing\_rate$ is the missing data rate for the
associated features and $\bigtriangleup Entropy$ is the information entropy change 
of the associated features, as computed before and after applying missing data handling.
$Accuracy$ is the classification accuracy, and the $\alpha$ and
$\beta$ are hyber-parameters to adjust the importance of 
the $Missing\_rate$ and $\bigtriangleup Entropy$.

The implications of using MAA and EAA are to adjust the classification accuracy based on
the magnitude of data variance, represented by the missing data rate and 
information entropy change for MAA and EAA, respectively. Instead of estimating 
the actual value of the data variance, we estimate the magnitude of data variance 
based on two assumptions: 1) features with a lower $Missing\_rate$ will
experience less data variance after applying missing data handling, and 2) a 
small $\bigtriangleup Entropy$ preserves the information entropy, leading to 
lower data variance. MAA uses the $Missing\_rate$ to select features with a 
lower missing data rate. Based on equation~\ref{maa}, whenever the missing 
data rate is high, the associated weight value shrinks, reducing the overall 
contribution of the corresponding feature. Likewise, EAA uses the $\bigtriangleup Entropy$, 
following the same procedure.

MAA is a straightforward method to select the features that exhibit low data variance. 
However, MAA also potentially eliminates important features that have a
relatively large missing data rate, but small $\bigtriangleup Entropy$ values. 
EAA, on the other hand, focuses on using the change in information entropy 
to adjust the classification accuracy, potentially being able to select important features MAA ignores.

\section{Experiments}\label{experiments}

\subsection{Dataset Specifications}


We generate a series of
multi-dimensional datasets, selected based on different data completion rate
thresholds (50\% to 80\%, in steps of 10\%), with participants (i.e., mothers) and features 
selected across all visits (V1 to V4). 

Table~\ref{tab:data} lists the details of the datasets. As indicated in the
table, the dataset is highly imbalanced in terms of the outcome class
distribution. Almost 90\% of the samples are negative (i.e., term births) and
only 10\% are positive (i.e., preterm birth).

\begin{table}[t!] 
\caption{Details of data.} 
\label{tab:data} \centering
\resizebox{\linewidth}{!}{ \begin{tabular}{ccccc} \hline
&\multicolumn{4}{c}{Data completion rate threshold}\\
&50\%&60\%&70\%&80\%\\
\hline Number of samples & 667 & 647 & 626 & 598\\ 
Number
of positive samples & 66 & 58 & 53 & 51\\ 
Number of negative samples & 601 & 589
& 573 & 547\\ 
Number of features & 1011 & 975 & 954 & 928\\ \hline 
\end{tabular}
} 
\end{table}

\subsection{Experimental Setup}

Table~\ref{tab:parameters} 
below lists the parameters and elements used to construct
the undersampling ensemble learning model.

\begin{table}[t!] 
\caption{Parameters and elements for the undersampling
ensemble learning model.} 
\label{tab:parameters} \centering
\resizebox{\linewidth}{!}{ \begin{tabular}{ccccc} \hline 
Decision Tree &
CART~\cite{duda}\\ 
\multirow{2}{*}{Ensemble Feature Selection Methods} & CLA,
WMA, OFA, CAA\\ 
&MAA, EAA ($\alpha:1, 0.5; \beta:2$)\\ 
Performance metrics & Accuracy,
AUC\\ 
Number of ensemble ways & 91\\ 
Number of selected features & 20\\ \hline
\end{tabular} } 
\end{table}

As indicated in table~\ref{tab:parameters}, we use CART from
sklearn~\cite{sklearn} as the Decision Tree model in our proposed undersampling
ensemble learning model. We leverage and compare CLA, WMA, OFA and CAA,
as well as
our proposed MAA and EAA, and evaluate accuracy and compute the AUC. 
We use an odd number as
the ensemble way size so that we can easily apply
majority voting during the performance evaluation
step shown in figure~\ref{fig:ensemblefs}. In addition, we found that an
ensemble way size
larger than 91 produces negligible performance gains in 
our experiments. Therefore, we
select 91 as the ensemble way size. We use the number of selected features
reported in previous studies that analyzed this same
dataset~\cite{dongbigdata} 
to provide a fair comparison. We use 10-fold cross validation in the performance evaluation step.


\section{Results}\label{results}
In this section, we present results from our experiments. We use AUC and
classification accuracy as the major measures of model
performance.

Given the same experimental setup as used in 
a previous study using the same dataset~\cite{dongbigdata}
(where only data from visits V1 and V2 were evaluated), 
the undersampling ensemble model combined with CAA produces a 
30\% improvement in AUC over a hybrid approach, while achieving an accuracy of
72\%. When using the dataset that combines all 4
visits, the base line method~\cite{dongbigdata} is equivalent
to random guessing (the number of samples is too small to apply 
the method explored in previous work~\cite{dongbigdata}). Given
this behavior, we exclude this option in our results. Using undersampling with
ensemble learning, on the other hand, overcomes this issue.

Tables~\ref{tab:performance1} and~\ref{tab:performance2} list the AUC and accuracy of CLA, WMA, OFA and CAA, while using both mean-based and similarity-based missing data imputation methods. From the results, we find that mean-based missing data imputation produces better performance across all ensemble feature selection methods. We see that CAA achieves the best performance in terms of both accuracy and AUC. Considering
that the hybrid approach presented in prior work~\cite{dongbigdata} 
produced results close to random guessing (AUC was close to 50\%), our proposed analytical model that includes CAA can improve AUC by 42\%, while achieving a 79\% accuracy, as seen in~Table~\ref{tab:performance1}.

\begin{table}[t!] 
\caption{Performance of CLA, WMA, OFA and CAA, with mean-based missing data imputation.}
\label{tab:performance1} \centering \resizebox{0.8\linewidth}{!}{
\begin{tabular}{ccccc} \hline 
& CLA & WMA & OFA & CAA\\
\hline
Accuracy & 0.72 & 0.75 & 0.76 & \textbf{0.79}\\
AUC & 0.68 & 0.61 & 0.68 & \textbf{0.71}\\
\hline
\end{tabular} } 
\end{table}

\begin{table}[t!] 
\caption{Performance of CLA, WMA, OFA and CAA, with similarity-based missing data imputation.}
\label{tab:performance2} \centering \resizebox{0.8\linewidth}{!}{
\begin{tabular}{ccccc} \hline 
& CLA & WMA & OFA & CAA\\
\hline
Accuracy & 0.41 & 0.38 & 0.73 & \textbf{0.74}\\ 
AUC & 0.57 & 0.52 & 0.69 & \textbf{0.71}\\
\hline
\end{tabular} } 
\end{table}

Next, we present the results based on data selected across different data
completion rates. As CLA and WMA do not have as 
good performance as compared to OFA
and CAA, we only present the results from OFA, CAA, MAA and EAA, with mean-based missing data imputation.

\begin{figure}[t!] \centering 
\includegraphics[width=.9\columnwidth]{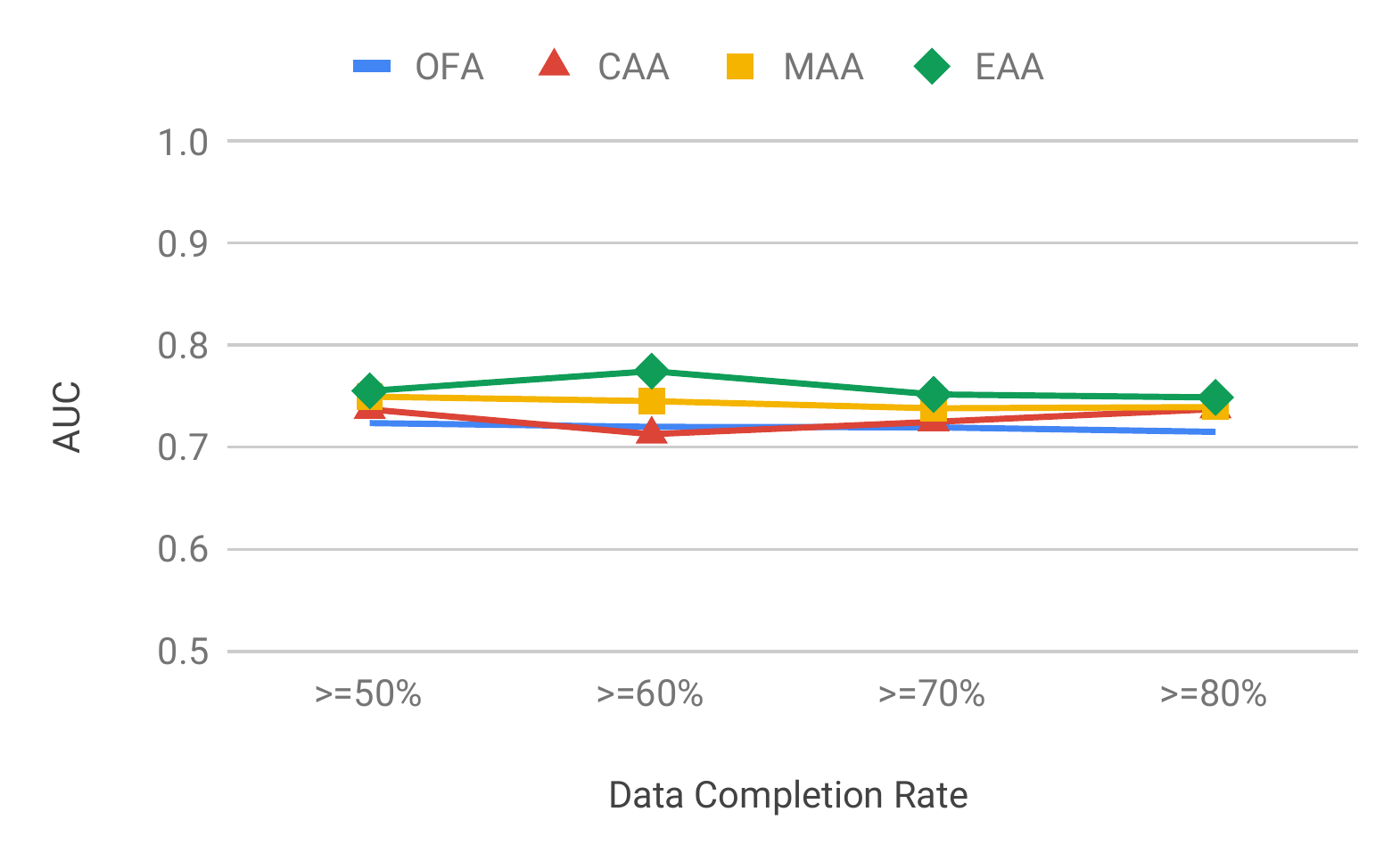}
\caption{AUC of OFA, CAA, MAA and EAA.} 
\label{fig:auc}
\end{figure}

\begin{figure}[t!] \centering 
\includegraphics[width=.9\columnwidth]{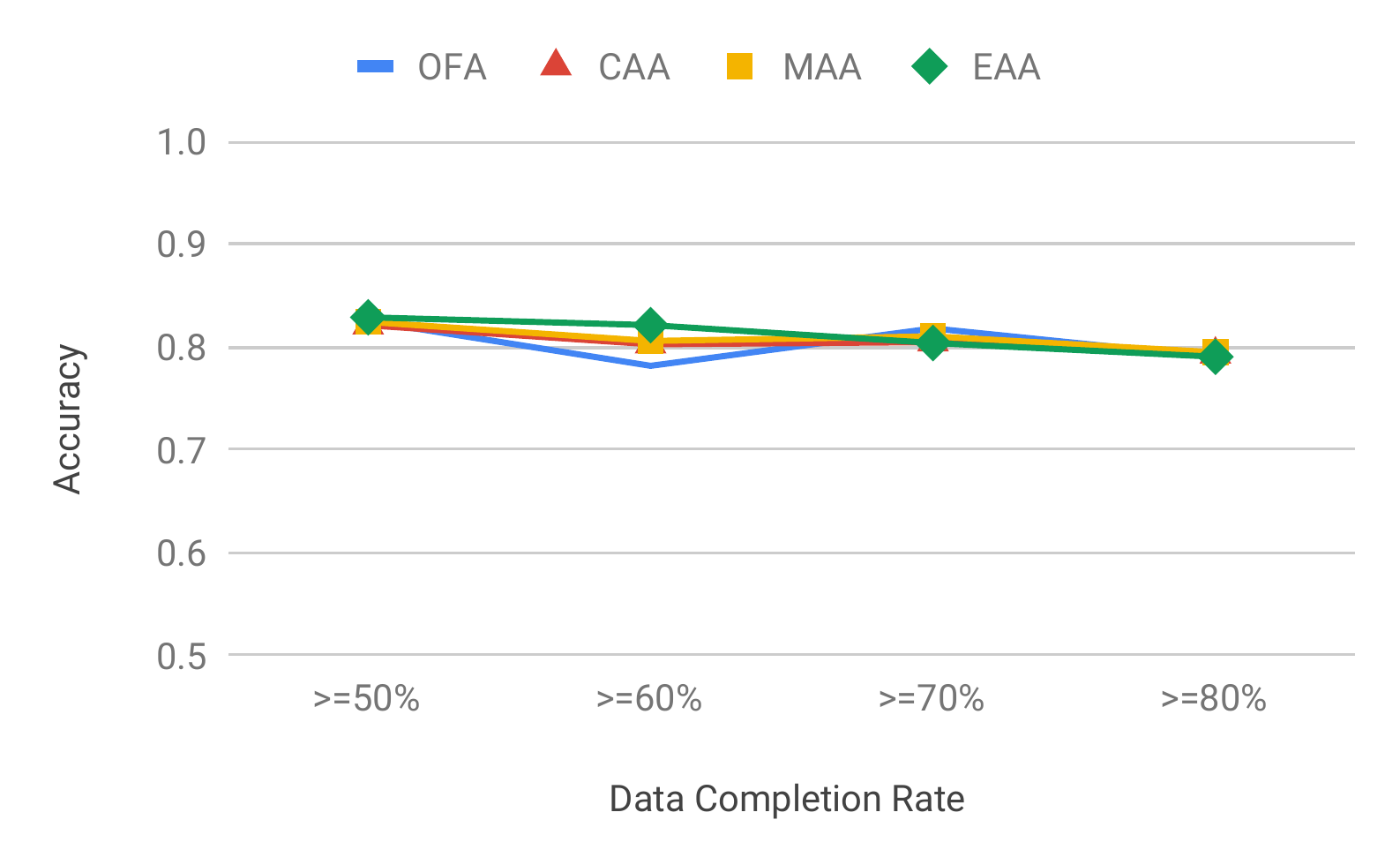}
\caption{Accuracy of OFA, CAA, MAA and EAA.} 
\label{fig:accuracy} 
\end{figure}

\begin{figure}[t!] \centering
\includegraphics[width=.9\columnwidth]{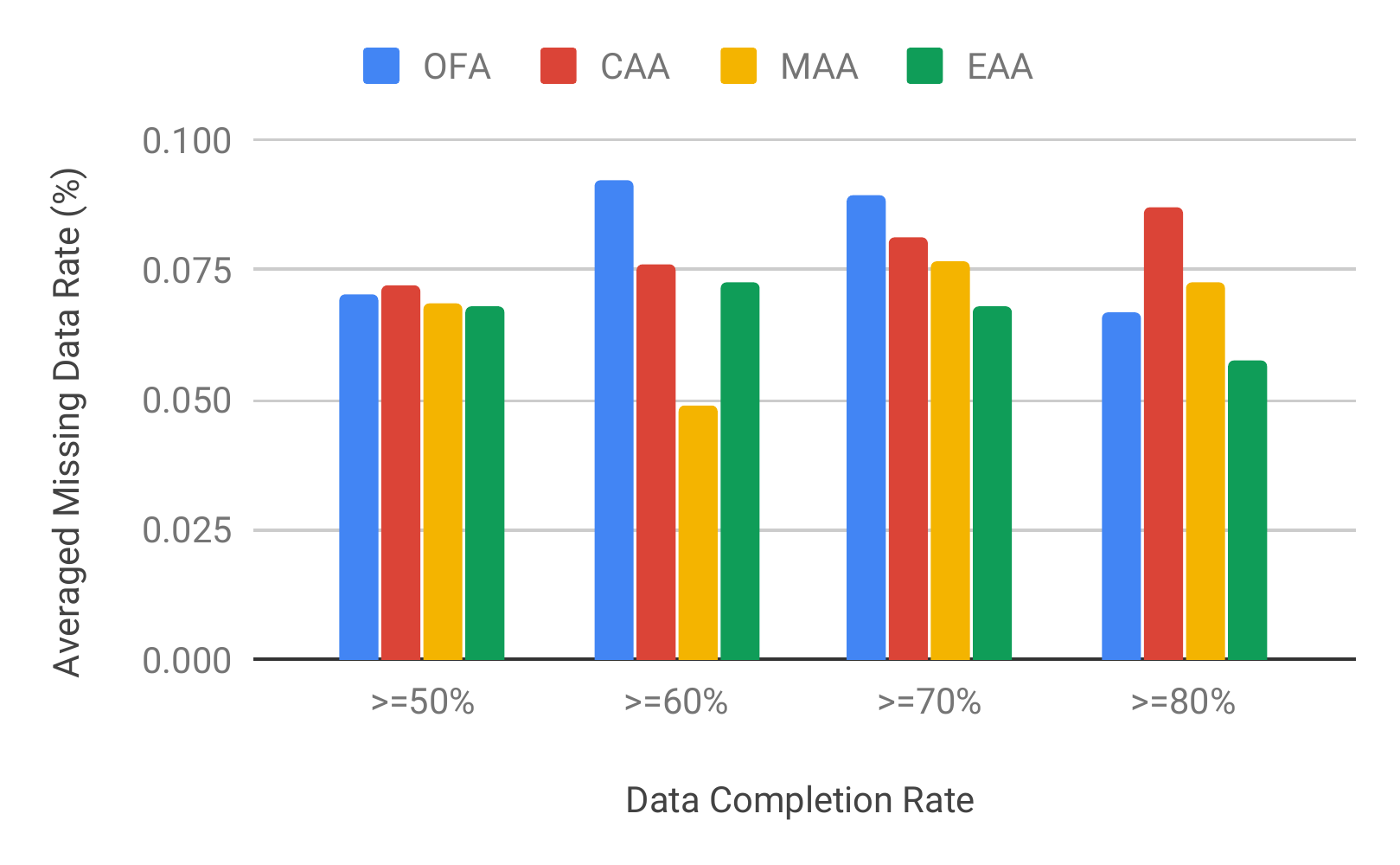} 
\caption{Averaged missing
data rate of features selected by OFA, CAA, MAA and EAA.}
\label{fig:missingdatarate}
\end{figure}

\begin{figure}[t!] \centering
	\includegraphics[width=.9\columnwidth]{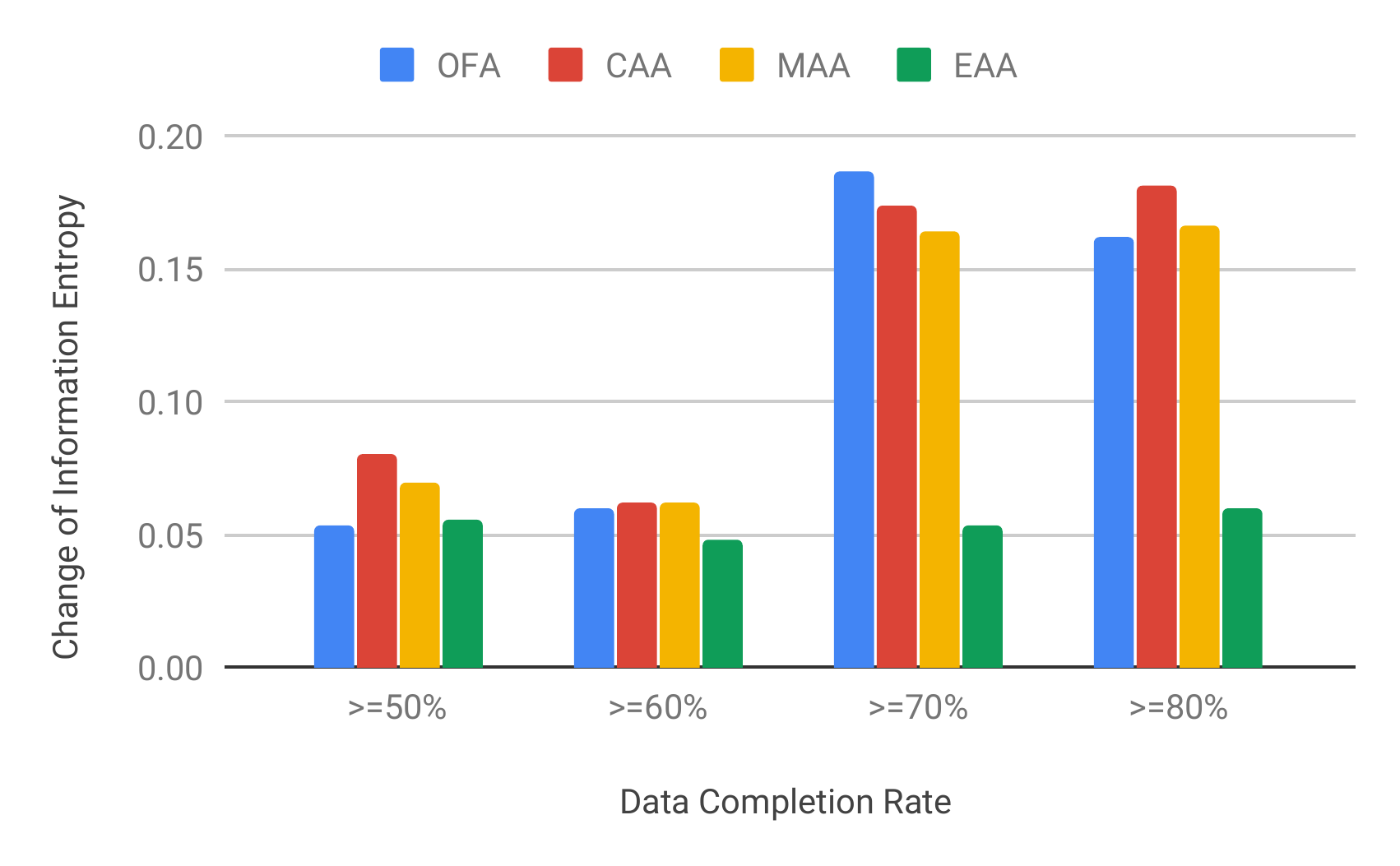} 
	\caption{Averaged information entropy change of features selected by OFA, CAA, MAA and EAA.}
	\label{fig:entroychange}
\end{figure}

Figures~\ref{fig:auc} and~\ref{fig:accuracy} show the AUC and accuracy
with ensemble feature selection with OFA, CAA, MAA and EAA, while varying the data
completion rate from $>=50\%$ to $>=80\%$, in steps of 10\%. From 
these results we can see that EAA has the best performance. Both MAA and EAA have slightly better performance than OFA and CAA. They are both effective in selecting features, while improving evaluation performance.
Figures~\ref{fig:missingdatarate} and~\ref{fig:entroychange} 
show the average missing data rate and the change in information entropy of
features selected by OFA, CAA, MAA and EAA. From Figure~\ref{fig:missingdatarate}, the feature set
selected by MAA presents a lower missing data rate as compared to OFA and CAA. This is expected, as MAA adjusts the classification 
accuracy using the missing data rate. EAA, surprisingly, performs more effectively in selecting features with a lower missing data rate, especially when the data completion rate is larger than 70\%. The feature set selected by EAA, as expected, results in a lower $\bigtriangleup Entropy$, as shown
in Figure~\ref{fig:entroychange}. From the results, MAA and EAA are found
to be more effective at providing low-variance feature selection, as measured
using the missing data rate and the $\bigtriangleup Entropy$, respectively.





\section{Conclusion}\label{conclusion}
In this paper, we present an analytical model for identifying potential
key factors contributing to the high rates of preterm birth in 
northern Puerto Rico, based on the
dataset available from the NIEHS P42 
Center. We highlight the challenges of
analyzing a diverse dataset with class imbalance in 
the target variable, as well as a
high missing rate in predictors. 

To address these issues, we
propose an undersampling ensemble learning approach leveraging many 
state-of-the-art ensemble feature selection methods, including CLA, WMA, OFA and CAA. 
We found that a undersampling ensemble model equipped with CAA can
achieve a 42\% improvement in AUC as compared to previous studies. We also propose two novel
feature selection methods, MAA and EAA, limiting the 
variance introduced by missing data handling. With OFA and CAA as a baseline, 
we evaluate MAA and EAA.  Our results show that MAA and EAA are effective at selecting features with lower data variance, while providing slightly better performance than CAA.

\ifCLASSOPTIONcompsoc
\else
 \section*{Acknowledgment} The work presented in this paper is supported in part by NIEHS P42 Program award P42ES017198, and NSF awards OAC-1559894 and IIS-1546428.
\fi



%
\balance

\bibliographystyle{IEEEtran}                                        
\bibliography{IEEEfull}

\end{document}